\documentclass[prl,twocolumn,amsfonts]{revtex4}
\usepackage{amsfonts}
\usepackage{amsmath}
\usepackage{bm}
\usepackage{graphicx}
\usepackage{epsf}

\def\prd{ Phys. Rev. D }
\def\pra{ Phys. Rev. A }
\def\prl{ Phys. Rev. Lett. }

\def\pla{ Phys. Lett. A }
\def\jmp{ J. Math. Phys. }
\def\rmp{ Rev. Mod. Phys. }
\def\etal{{\it et al.\/}}

\newcommand{\la}{\langle}
\newcommand{\ra}{\rangle}
\newcommand{\up}{\uparrow}
\newcommand{\down}{\downarrow}

\newcommand{\beq}{\begin{equation}}
\newcommand{\eeq}{\end{equation}}
\newcommand{\beqa}{\begin{eqnarray}}
\newcommand{\eeqa}{\end{eqnarray}}

\begin{document}

%%%%%%%%%%%%%%%%%%%%%%%%%%%%%%%%%%%%%%%%%%%%%%%%%%%%%%%%%%%%%%%%%%%%%%%%
%%%%%%%%%%%%%%%%%%%% Local Definitions %%%%%%%%%%%%%%%%%%%%%%%%%%%%%%%%%

\newcommand{\suba}[1]{_{_{A_{#1}}}}
\newcommand{\subat}[1]{_{_{\widetilde{A}_{#1}}}}
\newcommand{\subb}[1]{_{_{B_{#1}}}}
\newcommand{\subbt}[1]{_{_{\widetilde{B}_{#1}}}}
\newcommand{\subab}[1]{_{_{A_{#1}B_{#1}}}}
\newcommand{\subabt}[1]{_{_{\widetilde{A}_{#1}\widetilde{B}_{#1}}}}

%%%%%%%%%%%%%%%%%%%%%%%%%%%%%%%%%%%%%%%%%%%%%%%%%%%%%%%%%%%%%%%%%%%%%%%%
%%%%%%%%%%%%%%%%%%%%%%%%%%%%%%%%%%%%%%%%%%%%%%%%%%%%%%%%%%%%%%%%%%%%%%%%
\title{Violating Bell's inequalities in the vacuum}

\author{ Benni Reznik }
\author{ Alex Retzker }
\author{ Jonathan Silman }
\affiliation{ School of Physics and Astronomy,
Tel-Aviv University, Tel Aviv 69978, Israel. }
\date{October 9, 2003}
\begin{abstract}
\bigskip
We employ an approach wherein vacuum entanglement is directly
probed in a controlled manner. The approach consists of having a pair of initially
nonentangled detectors locally interact with the field  for
a finite duration, such that the two detectors remain causally
disconnected, and then analyzing the resulting detector mixed state.
 It is demonstrated
that the correlations between arbitrarily far-apart regions of the vacuum of a relativistic free scalar field
cannot be reproduced by a local hidden-variable model,
 and that as a function of the distance $L$ between the
regions, the entanglement decreases at a slower rate than $\sim
exp(-(L/cT)^3)$.
\end{abstract}
%\pacs{PACS number(s) 03.65.Bz}%
\maketitle

The vacuum state of a relativistic free field is
entangled. For two complementary regions of spacetime, such as
$x<0$ and $x>0$, this entanglement is closely related to the Unruh
acceleration radiation effect \cite{unruh1976}, and gives rise to
a violation of Bell's inequalities \cite{bell1964,summers1985}. For two
fully separated regions, entanglement persists
\cite{halvorson2000}, although, it is not known
whether Bell's inequalities are violated, or how entanglement
decays with the increase of
separation, as compared to correlations. Similar questions concerning
 entanglement have been
 addressed in the case of discrete models
 \cite{osborne2002, osterloh2002, vidal2003}.

\begin{figure}
{\vspace{1.3em}
\includegraphics[width=5cm,keepaspectratio]{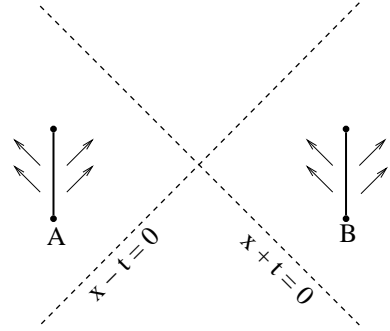}}\hspace{1cm}
\caption{ The world lines of detectors {\em A} and {\em B} are
shown for the duration of the interaction. The
horizontal and vertical axes are space and time respectively. The
arrows denote the emitted radiation. Notice that the radiation
emitted by detector $A$ ($B$)  does not affect detector $B$ ($A$),
since for $t>T$ the interaction is switched-off.
 } \label{purify}
\end{figure}

In this letter we shall study this problem by probing the
field's entanglement with a pair of localized two-level detectors
\cite{reznik2003}. This is done as follows. A state is prepared in which the two detectors
are not entangled with one another, or the field. We then have
 each of the detectors is made to locally interact with the field
for a finite duration, such that the detectors remain causally
disconnected throughout the process (Fig. \ref{purify}). Since
entanglement cannot be produced locally \cite{bennett1996}, once the interaction  is over, the
net entanglement between the detectors  must necessarily have its origin in vacuum
correlations. The interaction thus serves as a means of
redistributing entanglement between the field and the detectors.
We shall show that for arbitrarily far-apart regions, the detectors'
 final mixed state, after filtering, violates Bell's inequalities,
 and in the process obtain a lower bound on the amount of vacuum entanglement.

To set-up the model, we shall assume that the detectors are
localized within a region of a typical scale of $R$,
 and are separated by a much
larger distance $L>>R$. Consistency with relativity requires
us to use detectors of a rest-mass $M$, for which
$R>>\lambda_{Compton}=\frac{\hbar}{Mc}$. In this limit, the effects of
 both detector pair-creation, and the ``leakage'' of each detector's wavefunction to the
outside of its localization region, become exponentially small, of the
order of $\sim \exp(-\frac{2cMR}{\hbar})$ \cite{hegerfeldt, peresRMP}.
 Note, that this ensures that the overlap between the detectors' wavefunctions is negligible.
 Under these conditions, in their rest frame, the detectors can be described as nonrelativistic
quantum mechanical systems. Finally, we shall assume that, by
 means of an external coupler, each
detector's degrees of freedom can be coupled ``at will'' to
the field. Since the coupler need
not be of the same type as the studied field, we shall make the additional
assumption that it can
be described classically, and therefore does not generate
entanglement.

%(Similarly, the field which accelerates the detector in the Unruh
%effect has noting to do with the effect of thermalization.)

There have been several proposals for detector models which can
satisfy the above requirements; notably, the Unruh-Wald ``particle
in a box'' detector \cite{wald}, and the  DeWitt monopole detector
model \cite{dewitt}. In both models the detector Hamiltonian is $\frac{\Omega}{2} \sigma_z$, with  $\Omega$ being the
energy gap between the two levels and $\sigma_z$ a Pauli
matrix. The field-detector interaction Hamiltonian is
 \beq\label{zero}
  H_{int} =\epsilon(t)\int d^3x \psi(\vec x)
 (e^{+i\Omega t}\sigma^+ +e^{-i\Omega t} \sigma^-) \phi(\vec x,t)\, .
 \eeq
$\phi(\vec x,t)$ is a relativistic free scalar field in three
spatial dimensions, the $\sigma^\pm$ are the detector's energy raising
and lowering operators, and $\epsilon(t)$ governs the strength and
duration of the interaction. The function $\psi(\vec x)$ is a function of the detector's
spatial degrees of freedom, and is determined by the
model employed \cite{themodels,iontrap}.

Consider now a pair of DeWitt monopole detectors, $A$ and $B$,
that are localized about the coordinates $\vec x_A$ and $\vec x_B$,
respectively. These detectors interact with the field through
$H_{int}=H_{A}+H_{B}$, where $H_{A}$ and $H_{B}$ are interaction
Hamiltonians of the form of Eq. (\ref{zero}). The window
functions $\epsilon_A(t)$ and $\epsilon_B(t)$ are chosen to vanish
except for a finite duration $T$, such that $cT<< L=|\vec x_B -
\vec x_A|$, ensuring that the detectors remain causally
disconnected throughout the interaction.  In the following we
shall work in the Dirac interaction representation and employ
``natural'' units ($\hbar=c=1$).

\begin{figure}
{\vspace{1.3em} \includegraphics
[width=7cm,keepaspectratio]{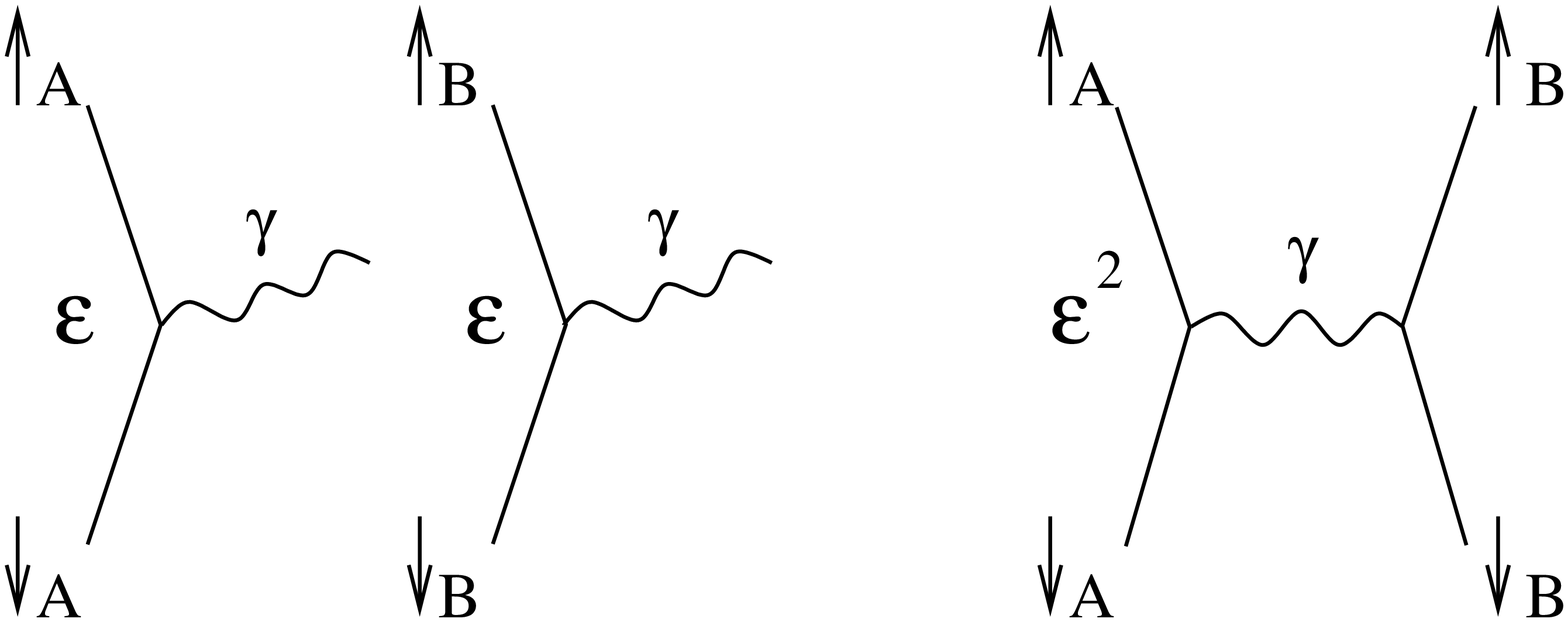}}\hspace{1cm}
\caption{Emission and Exchange Processes. The exchange amplitude,
portrayed on the right, is dominated by a single off-shell
emission followed by  an on-shell absorption, while the double
emission amplitude, portrayed on the left, consists of two
off-resonance processes. Thus, as $\Omega_{A,B}$ increases, the
emission term, $\Vert E_A\Vert \Vert E_B\Vert$, decreases more
rapidly than the exchange term, $|\la 0|X_{AB}\ra|$. }
\label{emission}
\end{figure}

Since the interaction takes place in two causally disconnected regions, the
Hamiltonians  $H_A$ and $H_B$ commute. The evolution operator $U$ for
 the whole system thus factors to a product of local unitary transformations
\beq
U
=\hat T[ e^{-i\int H_A(t) dt } \times e^{-i\int H_B(t') dt'}]\, ,
\eeq
where $\hat T$ denotes time ordering. This guarantees that $U$ does not change the net
entanglement between the regions.

We take the initial state of the detectors and the field to be $|\Psi_i\ra
=|\down_A\ra|\down_B\ra |0\ra$, where $|\down\ra$ and $|0\ra$
denote detector and field ground states respectively.
In the weak coupling limit $\epsilon_i(t)\ll1$ ($i=A,B$), expanding to the
 second order we get \beqa |\Psi_f\ra &=&\Bigl[({\bf 1} -
C)|\down\down\ra - \Phi^+_A\  \Phi^+_B|\up\up\ra
\nonumber \\
&-&i\Phi^+_A {\bf 1}_B|\up\down\ra - i{\bf 1}_A
\Phi^+_B|\down\up\ra \Bigr]|0\ra + O(\epsilon^3)\, , \eeqa where
$\Phi_i^\pm = \int dt\epsilon_i(t)e^{\pm i\Omega_i t}\int
d^3 x \psi_i({\vec x}) \phi({\vec x},t)$, and
$C=\frac{1}{2}\int\int dtdt'\hat{T}[H_A(t)H_A(t')]+
(A\leftrightarrow B)$. We observe that in the first term above,
the state of the detectors is unchanged, while in the second term
both detectors are excited and the final state of the field is
$|X_{AB}\ra\equiv \Phi^+_A\Phi^+_B|0\ra$. Since $|X_{AB}\ra$
contains either two photons or none, it describes, respectively,
an emission of a photon by each of the detectors or an exchange of
a single virtual-photon between them (Fig. \ref{emission}).
Finally, the last couple of terms describe an emission of a single
photon by either detector $A$ or  $B$. In this case the final
state of the field is $|E_A\ra\equiv
\Phi^+_A|0\ra$, or $|E_B\ra\equiv \Phi_B^+|0\ra$.

Tracing over the field degrees of freedom,
when working in the basis
$\{|i\ra \}_{i=0} ^3= \{\up\up,\up\down,\down\up ,\down\down\}$
and employing the notation $\Vert X_{AB}\Vert ^2 = \la X_{AB}|X_{AB}\ra$,
we obtain the detectors' reduced density matrix \cite{RDM}\\
\begin{equation}
\rho=
\left(
\begin{array}{cccc}
 \scriptstyle\Vert X_{AB}\Vert ^2       &      0         &      0       &
\scriptstyle - \la 0| X_{AB}\ra  \\
      0             &\scriptstyle \Vert E_{A}\Vert ^2    &\scriptstyle\la E_B|E_A\ra&
0\\
      0             &\scriptstyle\la E_A|E_B\ra  &  \scriptstyle\Vert E_B\Vert ^2   &
0 \\
 \scriptstyle - \la X_{AB}|0\ra &      0         &      0   & 1\scriptstyle - \Vert
E_{A}\Vert ^2-\Vert E_B\Vert ^2
\end{array} \right)\scriptstyle + \textit{O}(\epsilon^4)\, .
\label{density}
\end{equation}

Note the two types of off-diagonal terms.
The amplitude $\la 0| X_{AB}\ra$ acts to maintain coherence between
$|\down_A\down_B\ra$ and $|\up_A\up_B\ra$, while
the amplitude  $\la E_A|E_B\ra$
acts to maintain coherence
between $|\down_A\up_B\ra$ and $|\up_A\down_B\ra$.
It is the relative magnitude of these off-diagonal terms,
as compared
to the diagonal decoherence terms,
that determines whether the density matrix is entangled.

A density matrix is said to be inseparable or entangled iff it
cannot be expressed as a convex  sum of local density matrices
\cite{werner89}. In the present case of a $2\times2$ system, a
necessary \cite{peres1996} and sufficient \cite{horodecki1996}
condition for inseparability is that the negativity \cite{vidal01}
be positive. We shall therefore use the negativity as a measure of
entanglement. The following expression is obtained for the
negativity \beq\label{first} \mathcal{N}(\rho)\approx|\la
0|X_{AB}\ra| - \Vert E_A\Vert \Vert E_B\Vert >0\, . \eeq Physically
speaking, the inequality above is satisfied if the single
virtual-photon exchange process is more probable than the
off-resonance emissions of a single photon by each of the
detectors. The main contribution to the entanglement then arises
from states of the form \mbox{$\alpha|\down_A\down_B\ra +
\beta|\up_A\up_B\ra$}.

The inequality (Eq. \ref{first}) can be reexpressed as
\begin{widetext}
\begin{equation} \label{first1}
  \int_0^\infty \frac{d\omega}{L} \sin\left({\omega L}\right) e^{-{\omega}^2 R^2}
\tilde\epsilon_A
(\Omega_A+\omega)\tilde\epsilon_B(\Omega_B-\omega) >
\sqrt{\int_0^\infty \omega d\omega \, e^{-{\omega}^2 R^2} |\tilde\epsilon_A (\Omega_A+\omega)|^2
\int_0^\infty \omega d\omega \, e^{-{\omega}^2 R^2} |\tilde\epsilon_B (\Omega_B+\omega)|^2 }\, ,
\end{equation}
\end{widetext}
where the factor $e^{-w^2R^2}$ accounts for the ``smearing'' of the detectors, and $\tilde\epsilon_i(\omega)$
denotes the Fourier transform of $\epsilon_i(t)$.
(Note that for a massive field a factor of $\frac{w}{\sqrt{w^2+m^2}}$ must be added to each
integral.)
The term $\sin\left(\omega L\right)$ on the
left-hand side can be interpreted as an effective
window-function that characterizes the field's response
to the detectors.
When integrating over $\omega$, this function governs
the overall sign of each mode's contribution, and thus acts to
reduce the exchange amplitude. This destructive interference effect can
be minimized by employing a window-function
$\tilde\epsilon_A$ for which $ \sin(\omega L)\tilde\epsilon_A(\Omega_A+\omega)$
remains positive over a finite integration regime in the limit of large $L$.
A superoscillating function meets this requirement \cite{aharonov1988,berry}.
In particular the function \cite{benni1}
\begin{equation}
\tilde\epsilon_A(\omega) = f(\omega) J_0\Big(\sqrt{(\omega T)^2-N^2((L/T)^2-1)}\Big)\, ,
\end{equation}
where $f(\omega)$ is any function that converges  faster than
$1/\omega$ and has finite temporal support. We observe that
$\tilde\epsilon_A(\omega)$ is bounded in time as required, and
oscillates like $\sin(\omega L)$ about $\omega=\omega_{s}\pm
\sqrt{N}/2L$, where $T\omega_s=NL/T$, approximately $\sqrt N$
times, before gradually resuming ``normal'' slow oscillations for
larger values of $\omega$. The use of superoscillations, however,
is not without a price. For $\omega T< N\sqrt{(L/T)^2-1}$ the
function $\tilde\epsilon_A(\omega)$ decays exponentially,
rendering the exchange term, and hence the negativity,
exponentially small in $L$. The second window-function,
$\tilde\epsilon_B$, is a fixed hat function, convolved $k$ times
with itself. In $\omega$ space it assumes the form
$\Big(\frac{Sin(\omega T/k)}{\omega T/k}\Big)^k$. For given values
of $L$ and $T$ we choose the energy gaps $\Omega_A
=\frac{N}{T}\sqrt{(L/T)^2-1}$ and $\Omega_B = \omega_s-
\Omega_A\approx N/2L$. This choice of $\Omega_B$ fixes the center
of the window-function, $\tilde\epsilon_B(\Omega_B-\omega)$, in
the region of superoscillations. For large values of $L$, the
inequality (Eq. \ref{first1}) can then be approximated by
$\sqrt{N}\big(\frac{T}{L}\big)^2\frac{1}{\Omega_A T \Omega_B T}
\big(\frac{\tilde\epsilon_B(0)}{\tilde\epsilon_B(\Omega_B)}\big)>1$.
This ratio can be made arbitrarily large, at the expense of
reducing the negativity, by increasing $N$. To get a lower bound,
we take $N=(\frac{L}{T})^2$. Substituting the expressions for
$\Omega_A$ and $\Omega_B$ into the above approximation, it takes
on the form $(T/L)^5 (L/2T)^k >1$. For $k>5$ this ratio increases
with $L$. We thus get a lower bound on the negativity
 \begin{equation}
\mathcal{N}(\rho) \ge e^{- (L/cT)^3}\, .
 \end{equation}
Numerical computations show that this bound can be further improved.
Taking $N\ge L/T$, we get $\mathcal{N}(\rho) \ge e^{- (L/T)^2}$ \cite{invariance}.
The leakage of each detector's wavefunction to the outside of their localization regions
 introduces a correction of the order of $e^{-2MR}$ to the above expression.
 However, this correction can be made arbitrarily small by setting $e^{-(L/T)^2}>>e^{-2MR}$.
 Note that we are free to do this, since the mass scale, $M$, and the distance scale, $L$, are independent.

In passing, we would like to point out that in the case of the
electromagnetic field or any other free field,
 the analysis can easily be repeated.
Similar results are obtained in the case of the  finite duration
coupling of the detector's magnetic moment, or electric dipole to
the field. For a massive field, in the limit of large $L$, the
above result remains unchanged, because then the contribution to
the integrals arises from the range of frequencies $\omega\gg m$
for which the field effectively behaves as if massless.

\begin{figure}
{\vspace{1.3em}
\includegraphics[width=8cm,keepaspectratio]{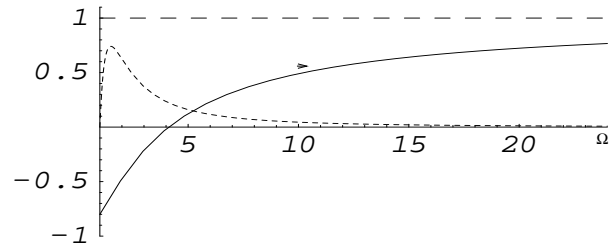}}\hspace{1cm}
\caption{Violation of the CHSH inequality. The dashed line
represents the negativity of the detectors before passing through
the filter, while the solid line represents the quantity
$M(\rho)-1$, which is calculated after the detectors have passed
through the filter. Since the CHSH inequality can be written in
the form $M(\rho)<1$  \cite{hor1}, we see that for
$\Omega\rightarrow\infty$,  the CHSH inequality is maximally
violated.}
\label{bell1}
\end{figure}

The reduced density matrix derived in the  previous section is
entangled. The question arises as to whether these vacuum correlations admit
 a local hidden-variable (LHV) description \cite{summers1985}.
Applying the Horodecki theorem \cite{hor1} to Eq. (\ref{density}),
we find that the detectors' final state does not violate the
CHSH  inequality \cite{clauser1969}. We shall now demonstrate that
by using local filters \cite{gisin1996}, a violation of the CHSH
inequality can be achieved for every separation distance, $L$.
Hence, the density matrix (Eq. \ref{density}) reveals a ``hidden''
nonlocality \cite{kwiat2001}, as in the case of Werner states
\cite{popescu1995}.

To show this we follow Gisin \cite{gisin1996}. Once the interaction with the
field has been switched-off we have each detector pass through the filter
\begin{equation}
f_{A,B}=\left(
\begin{tabular}{rc}
1 & 0  \\
0 & ${\eta}$ \\
\end{tabular} \right)\, .
\end{equation}
The density matrix is thus transformed according to
$\rho\to \rho_f=(f_A\otimes f_B) \rho (f_A\otimes f_B)$, so that the (nonnormalized)
filtered density matrix is given by
\begin{equation}
\left(
\begin{array}{cccc}
\scriptstyle \Vert X_{AB}\Vert ^2       &      0         &      0       &
\scriptstyle- \eta^2\la 0|X_{AB}\ra  \\
      0             &\scriptstyle \eta^2\Vert E_{A}\Vert ^2    &\scriptstyle\eta^2\la
E_B|E_A\ra&         0\\
      0             &\scriptstyle\eta^2\la E_A|E_B\ra  & \scriptstyle \eta^2\Vert
E_B\Vert ^2   &         0 \\
\scriptstyle  -\eta^2 \la X_{AB}|0\ra &      0         &      0       &\scriptstyle
\eta^4 (1-\Vert E_B\Vert ^2-\Vert E_{A}\Vert ^2)
\end{array} \right)
\label{densityf}
\end{equation}

Consider now the choice $\eta^2 \approx \la X_{AB}|0\ra$. We note that
the $00,\,33,\,03,\,30$ terms (Eq. \ref{densityf}) are now nearly equal, and of the
order of $|\la X_{AB}|0\ra|^2$. Ideally, in the absence of decoherence terms
in the inner block, these terms come close to reproducing a maximally
entangled state $\vert\up\up\ra -|\down\down\ra$.
Notice, however, that the decoherence terms are of the order of
$|\la X_{AB}|0\ra|
\Vert E_{A,B}\Vert^2$. Previously we have shown that
the ratio $|\la X_{AB}|0\ra|/(\Vert E_{A}\Vert \Vert E_{B}\Vert )$ can be made, by a
suitable
choice of window-functions,
arbitrarily large. Therefore in this extreme limit the relative
strength of the decoherence terms, as compared to the entangling terms,
is greatly reduced, and $\rho_f$ can be brought as close as we like
to a pure, maximally entangled state. This implies a maximal violation of Bell's
inequalities for the final
state of the detectors, and since initially the detectors are not
entangled (thus admitting a LHV description), it follows that correlations
between arbitrarily far-apart regions of the vacuum cannot be ascribed to a LHV model.

Maximal violation can be achieved at the price of reducing
the detectors' entanglement (negativity), which grows
smaller in the above limit (Fig. \ref{bell1}).
We now wish to quantify the more general case, for which the
final state is more entangled (larger negativity), but gives rise to
a weaker violation of Bell's inequalities.
Applying yet again the Horodecki theorem \cite{hor1}, we find that the CHSH inequality
is violated iff
\begin{equation}\label{cond1}
\frac{|\langle 0|X_{AB}\rangle|}{\Vert E_{A}\Vert \Vert E_{B}\Vert } > 4\frac{ \Vert
X_{AB}\Vert}
{|\la 0|X_{AB}\ra|}\, .
\end{equation}
Note that for-apart regions, this inequality is only slightly stronger than the condition for entanglement.

Interestingly,
if we repeat our process many times,
the resulting ensemble of entangled pairs can be reduced to a smaller one
of higher quality entangled pairs. This process is known
as distillation of entanglement, and is feasible for any inseparable
$2\times2$ mixed state \cite{horodecki1997}.
Furthermore, since Bell's inequalities are violated in our example, the two
detectors could be used directly for teleportation tasks, without having to
distill them first \cite{hor3}.

In conclusion, we have presented a new physical effect of vacuum
fluctuations which is associated with quantum nonlocality.
This effect stands in marked contrast to other vacuum phenomena, such as
the Lamb shift or the Casimir effect, which to some extent can be
``mimicked''  by classical stochastic local noise
\cite{milonni1994}.

\begin{acknowledgments}
We thank Y. Aharonov, L. Vaidman, S. Popescu, J. I. Cirac, I. Klich, and
A. Botero for helpful discussions and suggestions.
We acknowledge support from the ISF (Grant No. 62/01-1).
\end{acknowledgments}


\begin{thebibliography}{99}
\bibitem{unruh1976} W. G. Unruh, Phys. Rev. D {\bf 14}, 870 (1976).
\bibitem{bell1964} J. S. Bell, Physics, {\bf 1}, 195 (1964).
\bibitem{summers1985} S. J. Summers and R. Werner, Phys. Lett. A {\bf 110}, 257 (1985), and Commun. Math. Phys. {\bf 100}, 247, (1987).
\bibitem{halvorson2000} H. Halvorson and R. Clifton, \jmp {\bf 41}, 1711 (2000).
\bibitem{osborne2002} T. J. Osborne and M. A. Nielsen, \pra {\bf 66}, 032110 (2002).
\bibitem{osterloh2002} A. Osterloh, L. Amico, G. Falci, and R. Fazio, Nature {\bf 416}, 608 (2002).
\bibitem{vidal2003} G. Vidal, J. I. Latorre, E. Rico, and A. Kitaev, \prl
{\bf 90}, 227902 (2003).
\bibitem{reznik2003} B. Reznik, Found. Phys. {\bf 33}, 167 (2003).
\bibitem{bennett1996} C. H. Bennett \etal, Phys. Rev. Lett. {\bf 76}, 722 (1996).
\bibitem{hegerfeldt} G. C. Hegerfeldt, \prl {\bf 54}, 2395 (1985).
\bibitem{peresRMP} A. Peres and D. R. Terno \rmp {\bf 76}, 93 (2004).
\bibitem{wald}  W. G. Unruh and R. M. Wald, \prd {\bf 29}, 1047 (1984).
\bibitem{dewitt} B. S. DeWitt, in {\em General Relativity: An Einsten Centenary Survey}, edited by
S. W. Hawking, W. Israel, (Cambridge University Press, Cambridge, U.K., 1979).
\bibitem{themodels} In the Unruh-Wald model, the local interaction
$\delta(\vec x-\vec Q)\phi(x)$ is shown to give rise to a dipole-like
coupling, where the $\sigma^{\pm}$ denote raising and lowering between
``spatial'' eigenstates. In the DeWitt model, the field is coupled
to the detector's internal degrees of freedom. Note that we employ
 a trivially modified version of the DeWitt detector that accounts for its spatial ``smearing''.
\bibitem{iontrap} This Hamiltonian is reminiscent
of the standard phonon field-ion interaction employed in cavity
QED and ion-traps models: J. I. Cirac and P.
Zoller, \prl {\bf 74}, 4091 (1995). However, we do
not make the rotating wave approximation, in order for it to be
possible to identify vacuum entanglement.
\bibitem{RDM} In general, the reduced density matrix has 
components $\rho_{i=\pm 1,\, j=\pm 1}(\vec x_A, \vec x_B; \vec y_A, \vec y_B)$.
However, under the implicit assumption that the spatial and internal degrees
of freedom of each detector are initially not
entangled, and that the back-reaction on the spatial degrees of freedom 
can be ignored ($\Delta P \gg \Omega$),  the entanglement in the reduced 
density matrix, Eq. (4),
 will correspond to virtually all of the entanglement between the detectors. 
\bibitem{werner89} R. F. Werner, \pra {\bf 40}, 4277 (1989).
\bibitem{peres1996} A. Peres, \prl {\bf 77}, 1413 (1996).
\bibitem{horodecki1996} M. Horodecki, P. Horodecki, and R. Horodecki, Phys. Lett. A {\bf 223}, 1 (1996).
\bibitem{vidal01} G. Vidal and R. F. Werner, \pra {\bf 65}, 032314 (2002).
\bibitem{aharonov1988} Y. Aharonov, D. Z. Albert, and L. Vaidman, Phys. Rev. Lett. {\bf 60}, 1351 (1988).
\bibitem{berry} M. V. Berry, J. Phys. A: Math. Gen. {\bf 27}, L391 (1994).
\bibitem{benni1} B. Reznik, \prd {\bf 55}, 2152 (1997).
\bibitem{invariance} Under a Lorentz transformation the entanglement
 between the internal degrees of freedom may decrease
 (See [11] and references therein). However, the net entanglement between 
the detectors obviously remains unchanged when each 
detector's spatial degrees of freedom
 are taken into account. Our results are thus Lorentz invariant.
\bibitem{hor1} R. Horodecki, P. Horodecki, and M. Horodecki, \pla {\bf 200}, 340 (1995).
\bibitem{clauser1969} J. F. Clauser, M. A. Horne, A. Shimony, and R. A. Holt, \prl {\bf 23}, 880 (1969).
\bibitem{gisin1996} N. Gisin, \pla {\bf 210}, 151 (1996).
\bibitem{kwiat2001} P. G. Kwiat, S. Barraza-Lopez, A. Stefanov, and N. Gisin, Nature {\bf 409}, 1014 (2001).
\bibitem{popescu1995} S. Popescu, \prl {\bf 74}, 2619 (1995).
\bibitem{horodecki1997} M. Horodecki, P. Horodecki, and R. Horodecki, Phys. Rev. Lett. {\bf 78}, 574 (1997).
\bibitem{hor3} R. Horodecki, M. Horodecki, and P. Horodecki, \pla {\bf 222}, 21 (1996).
\bibitem{milonni1994} P. W. Milonni, {\em The Quantum Vacuum} (Academic Press, New York, 1994).
\end{thebibliography}
\end{document}